\documentclass[twocolumn,showpacs,preprintnumbers,prl,fleqn,superscriptaddress]{revtex4}
\usepackage{graphicx}
\usepackage{dcolumn}
\usepackage{bm}

\begin{document}

\title{Tuning the electron-phonon coupling in multilayer graphene with magnetic fields}
\author{C. Faugeras}
\affiliation{LNCMI-CNRS, BP 166, 38042 Grenoble Cedex 9, France.}
\author{M. Amado}
\affiliation{LNCMI-CNRS, BP 166, 38042 Grenoble Cedex 9, France.}
\affiliation{QNS-GISC. Departamento de F\'{\i}sica de Materiales,
Universidad Complutense, E-28040 Madrid, Spain}
\author{P. Kossacki}
\affiliation{LNCMI-CNRS, BP 166, 38042 Grenoble Cedex 9, France.}
\affiliation{Institute of Experimental Physics, University of
Warsaw, Poland}
\author{M. Orlita}
\affiliation{LNCMI-CNRS, BP 166, 38042 Grenoble Cedex 9, France.}
\author{M. Sprinkle}
\affiliation{School of Physics, Georgia Institute of Technology,
Atlanta, Georgia 30332, USA}
\author{C. Berger}
\affiliation{School of Physics, Georgia Institute of Technology,
Atlanta, Georgia 30332, USA} \affiliation{CNRS-Institut
N$\acute{e}$el, BP 166, 38042 Grenoble Cedex 9, France.}
\author{W.A. de Heer}
\affiliation{School of Physics, Georgia Institute of Technology,
Atlanta, Georgia 30332, USA}
\author{M. Potemski}
\affiliation{LNCMI-CNRS, BP 166, 38042 Grenoble Cedex 9, France.}
\date{\today }

\begin{abstract}
Magneto Raman scattering study of the E$_{2g}$ optical phonons in
multi-layer epitaxial graphene grown on a carbon face of SiC are
presented. At 4.2K in magnetic field up to 33 T, we observe a
series of well pronounced avoided crossings each time the
optically active inter Landau level transition is tuned in
resonance with the E$_{2g}$ phonon excitation (at 196 meV). The
width of the phonon Raman scattering response also shows
pronounced variations and is enhanced in conditions of resonance.
The experimental results are well reproduced by a model that gives
directly the strength of the electron-phonon interaction.
\end{abstract}
 \maketitle

Lattice vibrations in solids can be effectively modified via their
coupling to electronic excitations, as, for example, evidenced by
observations of Kohn anomalies in metals~\cite{Kohn59,Baron04}, of
coupled phonon-plasmon modes in polar
semiconductors~\cite{Mooradian66,Wysmolek06} or of different
phonon spectra in metallic and semiconducting
carbon-nanotubes~\cite{Reich04}. The electron-phonon interaction
is currently intensively studied in
graphene~\cite{Ando06,CastroNeto07,Yan07,Pisana07} which is a
two-dimensional crystal of carbon atoms arranged in a honeycomb
lattice and a semimetal with characteristic dispersions of
electronic states displaying Dirac cones near the Fermi
energy~\cite{Geim07,CastroNeto09}. The case of the long wavelength
optical E$_{2g}$-phonons at the $\Gamma$-point of the Brillouin
zone, which corresponds to the relative displacement of two
nonequivalent carbon atoms in the unit cell of
graphene~\cite{Ando06,CastroNeto07}, is of particular interest.
The perturbation due to this displacement is effective in inducing
the direct ($\Delta k = 0$) electronic transitions across the
Dirac point: E$_{2g}$-phonons efficiently couple to those low
energy interband excitations~\cite{Ando06,CastroNeto07} that are
unique to graphene. The spectrum of these excitations (its
$2E_{F}$ low energy onset) can be modified by tuning the Fermi
energy $E_{F}$. This was achieved in gated graphene flakes on
Si/SiO$_{2}$ substrates where the electrically modified
E$_{2g}$-phonon spectrum was traced with Raman scattering
methods~\cite{Yan07,Pisana07}.

The spectrum of the graphene E$_{2g}$-phonon is expected to be
even more severely modified by applying a magnetic field
perpendicular to the 2D plane~\cite{Ando07,Goerbig07}, i.e., when
a continuous spectrum of electronic excitations is transformed
into a series of quasi-discrete inter Landau level excitations
characteristic of a 2D system. In conditions of Landau
quantization, the electron phonon coupling has a resonant
character which is expected between the E$_{2g}$-phonon and
properly selected inter-Landau level excitations. The observation
of the effects of magneto-resonant electron-phonon coupling in
graphene structures, which we present in this letter, has been an
experimental challenge~\footnote[1]{The experimental approaches to
trace the magneto-phonon resonance effects in graphene flakes on
Si/SiO2 substrates have been presented by J. Yan et al. at the APS
2009 March meeting in Pittsburg and by S. Remi et al. at the KIAS
Graphene meeting in Seoul. Notably, S. Remi et al. has reported
the splitting and not a simple shift of the E$_{2g}$ phonon in the
Raman spectrum at B = 11 T of likely quasi-neutral graphene.}
aiming at verification of theoretical
predictions~\cite{Ando07,Goerbig07} and eventual confirmation of
the conclusions drawn from zero-field
measurements~\cite{Yan07,Pisana07}.

We report here on magneto-Raman scattering studies of the
E$_{2g}$-phonon band of multilayer epitaxial graphene on the
carbon face of a SiC substrate (MEG), in fields up to 33 T, and
low, liquid helium temperatures. The measured E$_{2g}$-band is
observed to be composed of two components: one fixed and another
oscillating with the applied field. The extracted oscillatory
component of the E$_{2g}$-band shows the theoretically anticipated
magneto-phonon multi resonance~\cite{Ando07,Goerbig07}. Pronounced
variations of the energy position and width of this line can be
understood in terms of a series of avoided crossings each time the
E$_{2g}$-phonon energy tends to match the energy of the specific
(defined by appropriate selection rules) inter Landau level
excitations. Comparison of the experimental results to existing
models allows us to analyze the characteristic strength of the
electron-phonon interaction in the investigated graphene system.
Possible origin of the field-independent component of the
E$_{2g}$-band is discussed.

\begin{figure}
\includegraphics[width=0.8\linewidth,angle=0,clip]{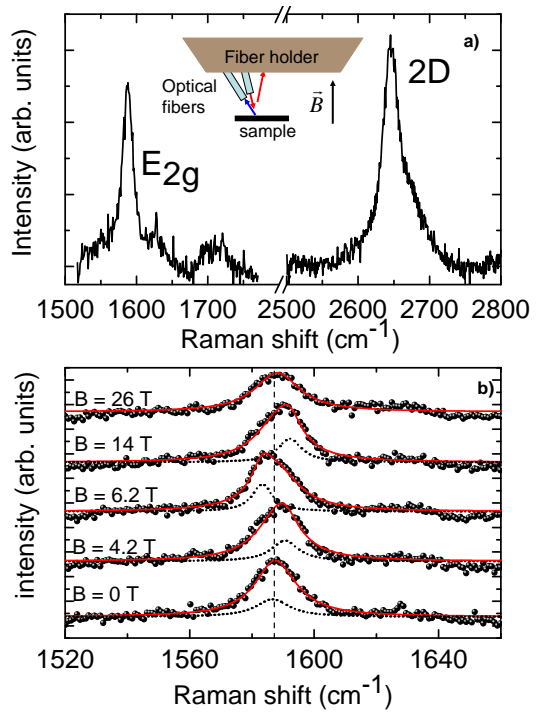}
\caption{\label{Fig1}(a) The E$_{2g}$ and 2D Raman scattering
bands measured at T$ = 4.2$~K and $\lambda$ = 720.7~nm laser
excitation at zero magnetic field. The scheme of the experimental
configuration is shown in the inset. (b) Background corrected
spectra of the E$_{2g}$ band, measured at different magnetic
fields. The solid red lines represent the fitted two-Lorentzian
component shape. One of this component is field independent and
fixed to solely reproduce the spectrum at B = 26 T. The
oscillatory with field component is shown with dotted black lines.
}
\end{figure}

Raman scattering spectra were measured using the Ti:Saphire laser
setup, tuned at accurately controlled wavelength in the range
$\sim\lambda$=720~nm, in order to minimize the superfluous Raman
signal of optical fibers which were used to transmit the
excitation light to, and to collect the Raman photons from the
sample. The sample was immersed in a helium gas kept at T$=4.2$~K
and placed in a resistive magnet delivering fields up to $33$~T.
The non-polarized Raman scattering spectra were measured in nearly
back-scattering Faraday geometry: the arrangement of the
excitation and collection fibers (both with core diameters of 200
$\mu$m) with respect to sample surface is sketched in the inset of
Fig.~\ref{Fig1}a. Typical excitation power was $\sim100$~mW,
spread over 600~$\mu$m -diameter laser spot on the sample. The
collected light was dispersed with a single grating spectrometer
(spectral resolution $\Delta\lambda$=0.3nm) equipped with nitrogen
cooled CCD detector and band pass filters were used to reject the
stray light.

The MEG structures which we have used for experiments were
fabricated by thermal decomposition of the carbon face of a
4H--SiC substrate~\cite{Berger04}. Two investigated structures,
both with a relatively large number of graphene layers, show
similar data and we present here the results obtained for the
sample which contains $\sim70$ layers. We used strongly
graphitized samples to enhance the Raman scattering signal,
particularly because the E$_{2g}$-phonon band in MEG samples
appears on the background of the two-phonon Raman scattering
processes in the SiC substrate. A number of experimental studies,
including magneto-absorption~\cite{Sadowski06,Orlita08} and
micro-Raman scattering measurements~\cite{Faugeras08}, show that
Dirac-like electronic bands, which are genuine of a single
graphene sheet, persist in MEG structures and that a large part of
these layers is practically neutral, with carrier densities as low
as $5\times10^{9}$ cm$^{-2}$ and fairly high mobilites up to
250000 cm$^2$/(V.s) \cite{Orlita08}. The electronic bands in MEG
grown on the carbon face of SiC are alike those of a single layer
because of preferentially rotational and not Bernal-type layer
stacking in this material \cite{Hass08,Lopes07,Latil07,Magaud09}.
Although we believe that the majority of the structure probed in
our experiments displays neutral graphene-like electronic system,
the MEG samples and in particular the strongly graphitized
specimens are not expected to be uniform on the scale of
600~$\mu$m defined by the actual size of the laser spot on the
sample. For example, the presence of Bernal stacked minority
inclusions in these structures has been seen with micro-Raman
scattering experiments~\cite{Faugeras08}.

The representative Raman scattering bands observed for our
$\sim70$ layers MEG sample in the absence of magnetic field are
shown in Fig.~\ref{Fig1}a. The so-called "2D" band, which is
frequently analyzed to characterize different graphene
structures~\cite{Ferrari06,Graf07}, appears at 2646~cm$^{-1}$ as a
~35 cm$^{-1}$-wide, slightly asymmetric line. A pure single
Lorentzian shape of the "2D" line is a signature of the electronic
system with simple Dirac-like electronic states, such as those
characteristic of graphene or of the majority part of the MEG
structures. We understand that the observation of the high energy
asymmetry in the "2D" line under the present experimental
conditions is a consequence of signal averaging from a relatively
large portion of the sample, which includes other than simple
graphene-like structures (i.e., Bernal stacked residues).

\begin{figure}
\includegraphics[width=0.9\linewidth,angle=0,clip]{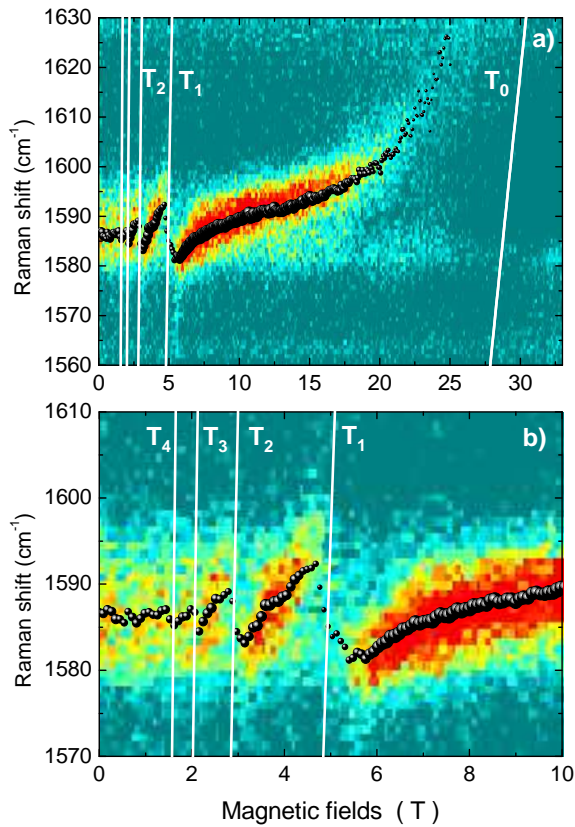}
\caption{\label{Fig2} a) Color map of the magneto-oscillatory
component of Raman scattering spectra of E$_{2g}$ band phonons as
a function of the magnetic field measured at T = 4.2 K under
$\lambda$=720.7~nm laser excitation. The extracted peak position
of this line is shown with full dots. Their size is proportional
to the line amplitude. Solid lines T$_{k}$ represent the energies
of the series of inter Landau level transitions:$L_{-k,(-k-1)}
\rightarrow L_{k+1,(k)}$, which couple to E$_{2g}$ phonon
($\tilde{c}=1.02\times10^{6} m.s^{-1}$ is assumed). b) Zoom on the
0 to 10 T range of magnetic fields.}
\end{figure}

The focus of this work is on the E$_{2g}$-phonon line which
clearly dominates over the background of two-phonon signal from
SiC substrate and peaks at 1586.5~cm$^{-1}$ in the Raman
scattering spectrum measured at zero magnetic field
(Fig.~\ref{Fig1}a). As shown in Fig.~\ref{Fig1}b with background
corrected spectra, the E$_{2g}$-phonon line is visibly affected by
the application of the magnetic field. A more detailed data
inspection shows however that the E$_{2g}$ line is practically
field independent at high field region above B$\gtrsim 24$~T.
Hence, one may anticipate that the observed magneto-Raman spectrum
of the E$_{2g}$-phonon is composed of two components: one
oscillating with field and the second one which is field
independent and dominant at B $> 24$~T. Indeed, as illustrated in
Fig.~\ref{Fig1}b, the E$_{2g}$ spectrum measured at any magnetic
field can be fairly well fit with two-Lorentzian functions (solid
red lines). The field independent component I$_{26} (h\nu)$ has
been set to reproduce the spectrum at B $= 26$~T. The resulting
oscillatory component is shown with dashed lines in
Fig.~\ref{Fig1}b. To present the full data collection, the I$_{26}
(h\nu)$ Lorentzian function has been extracted from each measured
spectrum and the resulting differential spectra (oscillatory part)
are shown in Fig.~\ref{Fig2}, in form of the intensity false color
map. Full points in this figure denote the peak position (center)
of the Lorentzian functions which fit the oscillatory component.

The magnetic field evolution of the spectra shown in
Fig.~\ref{Fig2} can be clearly seen as resulting from a series of
avoided crossings between the E$_{2g}$ phonon excitation and the
specific inter Landau level electronic transitions. The energy
ladder of Landau levels ($L_n$; $n=0,\pm 1, \pm 2, ...$) of the
two-dimensional Dirac-like electronic system is given by E$_n
=sgn(n) \tilde{c} \sqrt{2e \hbar B|n|}$, where the measured slope
of the Dirac cone (Fermi velocity) $\tilde{c}=1.02\cdot10^{6}m/s$
in the investigated structures~\cite{Sadowski06,Orlita08}. As
illustrated in Fig.~\ref{Fig2}, the E$_{2g}$-phonon couples to a
series of $L_n \rightarrow L_m$ transitions which fulfill the
$|n|-|m|= \pm 1$ selection rules and therefore appear at energies
$T_{k}=( \sqrt{k} + \sqrt{k+1} )E_{1}$ ($k=0,1,2, ..$). Notable,
the same selection rules apply to optically active inter Landau
level transitions which are observed in far-infrared
magneto-transmission
spectra~\cite{Ando07,Goerbig07,Sadowski06,Sadowski07}. The
amplitude of the observed "anticrossings" grows with the magnetic
field because, as a rule, the splitting ($\Delta$) between coupled
modes is proportional to the square root of the oscillator
strength of the uncoupled excitations and to the square root of
interaction parameter $\lambda$. In our case, these are electronic
excitations which become enhanced by the magnetic field. This is
primarly due to the increasing with B-field Landau level
occupation ($eB/h$ degeneracy). If broadening is neglected we
roughly expect $\Delta \sim \sqrt{\lambda B}$ and derive more
rigorously \cite{Goerbig07}: $\Delta =\sqrt{2 \lambda }E_1$, for
our neutral Dirac-like system.

\begin{figure}
\includegraphics[width=0.85\linewidth,angle=0,clip]{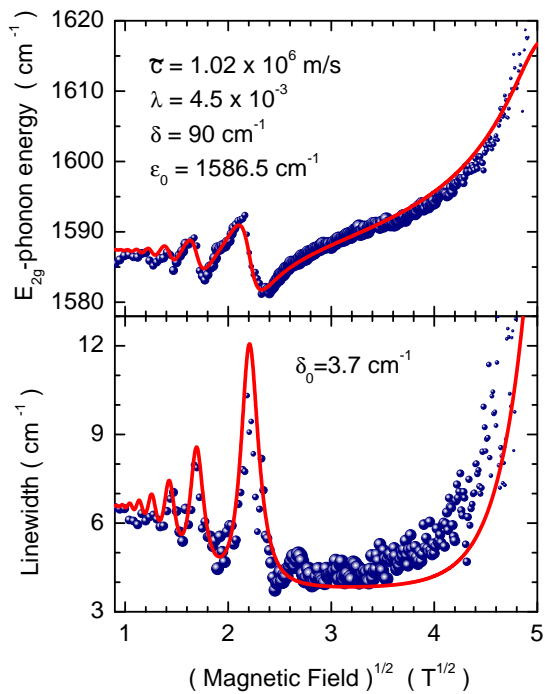}
\caption{\label{Fig3} Magnetic field evolution of the peak
position (upper part) and the linewidth (HWHM) of the
magneto-oscillatory component of the measured E$_{2g}$ Raman band.
Solid lines represent the result of modelling of the data (with
the parameters specified in the figure) along the procedure
described in the text.}
\end{figure}

For the sake of more quantitative data analysis we consider both
the characteristic phonon energy as well as its broadening which
is another inherent element of mode coupling effects. In
 Fig.~\ref{Fig3} we re-plot the center and add the extracted
linewidth (HWHM) of the Lorentzian peak which fits the
magneto-oscillatory component of the measured E$_{2g}$-line. The
magneto-oscillations in the peak position are accompanied by
oscillations in the linewidth which, as expected, is enhanced
under resonant conditions. To reproduce the data of
Fig.~\ref{Fig3}, we adopt the approach of Ando \cite{Ando07} to
our neutral Dirac-like system and derive the phonon energy
$\epsilon$ and broadening parameter $\Gamma$ by extracting
$\tilde{\epsilon}=\epsilon-i\Gamma$ from the equation which
defines the poles of the phonon Green's function:
$$\tilde{\epsilon}^{2}-\epsilon_{0}^{2} = 2 \epsilon_{0} \lambda E_{1}^{2} \sum _{k=0} ^{\infty} \{\frac{T_{k}}{(\tilde{\epsilon}+i\delta)^{2}-T_{k}^{2}}+\frac{1}{T_{k}}\}$$
where $\epsilon_{0}$ stands for the phonon energy of the neutral
system at B=0~T and $\delta$ accounts for the broadening
characteristic for electronic excitations. The measured linewidth
has been assumed as a convolution sum
$\sqrt{\delta_{0}^{2}+\Gamma^{2}}$, where $\delta_{0}$ accounts
for other, than electron-phonon coupling, broadening mechanisms.

To calculate the energy and broadening of the E$_{2g}-$ phonon
spectrum we fix the value of $\tilde{c} = 1.02\times10^{6} m/s$ as
derived from far-infrared magneto-absorption experiments on MEG
samples~\cite{Sadowski06,Orlita08}. Working effectively with only
two adjustable parameters $\delta$ and $\lambda$ ($\delta_{0}$ =
3.7~cm$^{-1}$ is the background correction to the linewidth
oscillations) the experimental data can be fairly well reproduced.
Optimal choice of parameters is: $\delta = 90$ cm$^{-1}$ and
$\lambda = 4.5\times10^{-3}$. The extracted strength of
electron-phonon coupling is in excellent agreement with
estimations of $\lambda$ in the range of $4-5.5\times10^{-3}$ from
the detailed analysis of the dispersions of the E$_{2g}$ in
graphene~\cite{Lazzeri06a} and/or measurements of electrically
tuned E$_{2g}$-phonon in graphene~\cite{Yan07,Pisana07}. We also
note that the width $\delta_{0}$ which is measured in between well
separated resonances, i.e., when the effects of electron-phonon
interaction are switched off, appears to be practically identical
to its value measured under corresponding conditions
($2E_{F}>\epsilon_{0}$) on graphene flakes on
Si/SiO$_2$~\cite{Yan07}. The parameter $\delta$ accounts for
broadening of electronic states and its actual value reasonable
matches the width of magneto-absorption transitions expected in
our sample at high magnetic fields~\cite{Plochocka08}. For
simplicity $\delta$ is assumed constant, thought its increase with
field ($\sim\sqrt{B}$) which is expected from magneto-absorption
experiments~\cite{Orlita08} could improve the data simulation
(larger broadening of the high field resonance).

The puzzle of the presented results is the origin of the field
independent component of the E$_{2g}$ spectrum which, we note,
represents the significant portion of the total spectral weight.
The Bernal-stacked inclusions are first possible candidates to
account for this contribution (we have checked that in fields up
to 14T the magneto-Raman spectrum of the E$_{2g}$ band of natural
graphite sample is practically field independent). However,
inspecting the shape of the measured 2D band, we can rule out that
the contribution from these inclusions is as large as 65$\%$
(extracted for the B = 0 T spectrum). Another candidate is a
signal from highly doped layers close to the interface. This is
also scarcely probable since such signal should be significantly
shifted in frequency from that of the majority neutral graphene
layers, which is not seen. We speculate the field independent
component of the E$_{2g}$ line is an inherent property of the
graphene-like systems~\footnotemark{} what remains to be
understood.

In conclusion, we have investigated the magnetic field evolution
of the Raman scattering response of the E$_{2g}$ phonons in a
graphene-like system in fields up to 33 T. A pronounced avoided
crossing behavior of the phonon energy and oscillations of the
line width are observed each time the optically active inter
Landau level excitation is tuned in resonance with the phonon
energy. The experimental results are well accounted for by
existing theoretical models~\cite{Ando07,Goerbig07} of
electron-phonon coupling in graphene, with the estimated
electron-phonon coupling constant $\lambda=4.5\times10^{-3}$.

\begin{acknowledgments}
We would like to acknowledge fruitful discussions with V. Fal'ko
and M.O. Goerbig. We acknowledge T. Kazimierczuk for help with the
data analysis and A. Wysmolek for his help in the initial stage of
this experiment. Part of this work has been supported by ANR
projects ANR-08-JCJC-0034-01, ANR-06-NANO-019 and by the French
Embassy (Partner University Fund). One of us (P.K.) is financially
supported by the EU under FP7, contract no. 221515 `MOCNA'. M. A.
thanks the Cariplo Foundation (project QUANTDEV), MICINN (Project
MOSAICO) and JCYL SA052A07 for support.

\end{acknowledgments}

\bigskip


\end{document}